\documentclass[journal,twoside,web]{ieeecolor}
\usepackage[hidelinks]{hyperref} % for hyperlinks, remove boxes

\usepackage{jsen}
\usepackage{cite}
\usepackage{amsmath,amssymb,amsfonts}
\usepackage{algorithmic}
\usepackage{graphicx}
\usepackage{textcomp}
\usepackage{wrapfig}

\usepackage{orcidlink}
\usepackage{main}
\usepackage{multirow}
\usepackage{bm}
\usepackage{bbold}
\usepackage{subfig}

\usepackage{url}
\usepackage{nicefrac}
\usepackage{hhline} % for \hhline
\usepackage{mathtools} % for :=
\usepackage{booktabs} % for better tables
\usepackage{xspace}
\usepackage{xfrac}

\usepackage{tikz}
\usepackage{pgfplots}
 
\usepackage{microtype} % for better typography, including better line breaks and spacing
\usepackage{etoolbox} % for \relpenalty and \binoppenalty
\AtBeginDocument{ % avoid breaking equations across pages 10000 = infinite penalty, 7000 = high penalty for breaking binary operators
    \relpenalty=10000
    \binoppenalty=7000
}

\pgfplotsset{compat=1.18}
\usetikzlibrary{arrows.meta, positioning, calc}
\usepgfplotslibrary{polar}

\newcommand{\mailto}[1]{\href{mailto:#1}{#1}}

\usepgfplotslibrary{external}
\def\isCI{}

\ifdefined\isCI
\else
\fi

\graphicspath{{./Figures/}}

\def\BibTeX{{\rm B\kern-.05em{\sc i\kern-.025em b}\kern-.08em
    T\kern-.1667em\lower.7ex\hbox{E}\kern-.125emX}}

\title{Utilizing Missed Detections in Directional Sensitivity-Based DOA Estimation}

\definecolor{abstractbg}{rgb}{0.89804,0.94510,0.83137}
\makeglossary
\usepackage{lipsum}

\begin{document}

\author{
    Gustav Zetterqvist \orcidlink{0000-0001-6672-4472}, %\IEEEmembership{Member, IEEE}, 
    Fredrik Gustafsson \orcidlink{0000-0003-3270-171X}, \IEEEmembership{Fellow, IEEE}, \\ 
    and Gustaf Hendeby \orcidlink{0000-0002-1971-4295}, \IEEEmembership{Senior Member, IEEE}
\thanks{This work was supported in part by the Wallenberg AI, Autonomous Systems and Software Program (WASP) through the Knut and Alice Wallenberg Foundation; in part by Excellence Center at Linköping-Lund in Information Technology (ELLIIT); and in part by Security Link. }
\thanks{The authors are with the Department of Electrical Engineering, Linköping University, 581 83 Linköping, Sweden (e-mail: \mailto{gustav.zetterqvist@liu.se}, \mailto{fredrik.gustafsson@liu.se}, \mailto{gustaf.hendeby@liu.se}). }}

\IEEEtitleabstractindextext{%
\fcolorbox{abstractbg}{abstractbg}{%
\begin{minipage}{\textwidth}%
\begin{wrapfigure}[12]{r}{3.15in}% no of lines, right-aligned, width
    %% See details https://journals.ieeeauthorcenter.ieee.org/create-your-ieee-journal-article/prepare-supplementary-materials/#graphicalabstract
\includegraphics[width=3in]{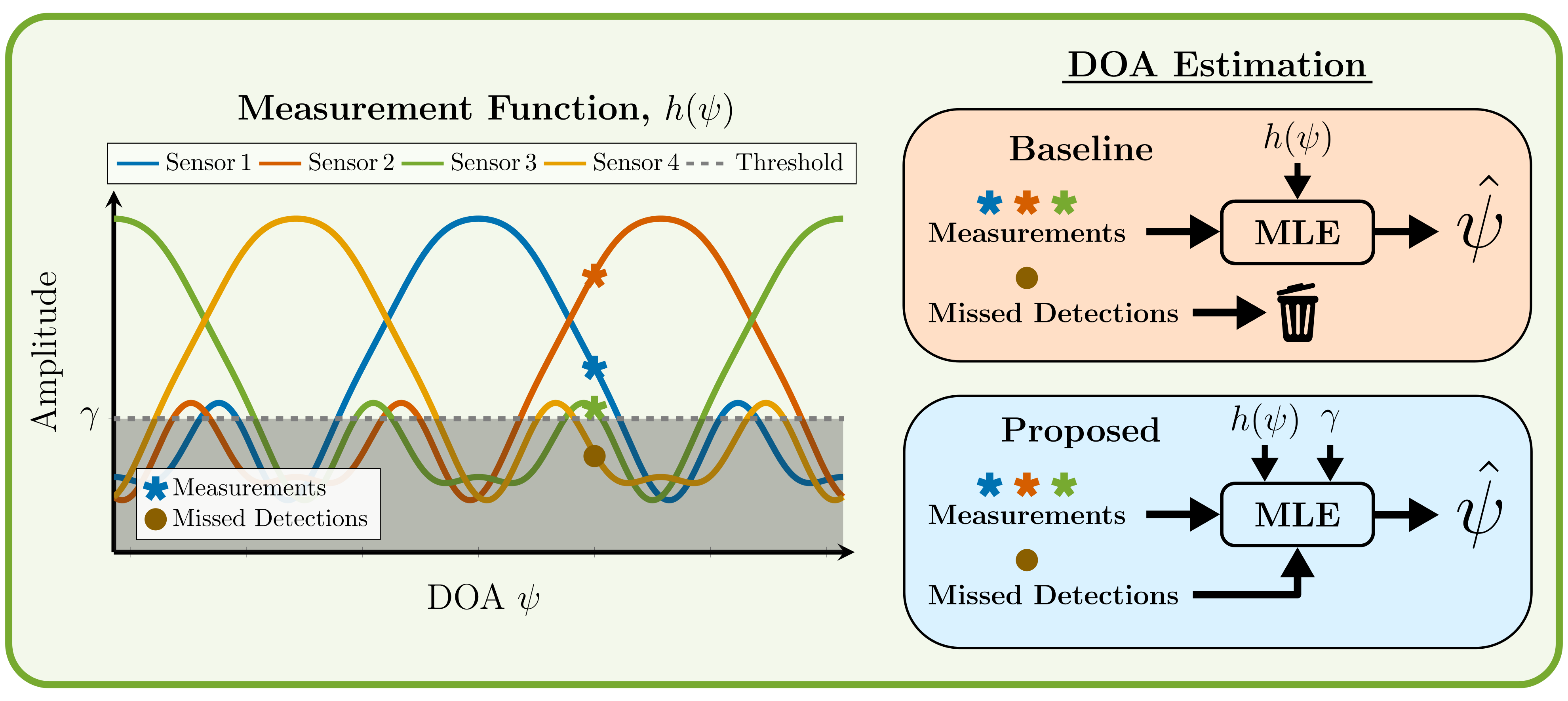}% Add graphical abstract here 
\end{wrapfigure}%
\begin{abstract}
    This paper introduces a signal strength-based \gls{doa} estimation approach for directional sensors that explicitly accounts for missed detections.
    In traditional phase-based \gls{doa} estimation frameworks, negative information from expected emitters that fall below the detection threshold lies outside the scope of standard measurement models.
    Unlike phase-based \gls{doa} estimation methods, the proposed approach relies only on received signal strength measurements. 
    As a result, missed detections arise naturally from the sensing and detection process and convey valuable information via the known detection thresholds.
    By incorporating both detected signals and missed detections into the likelihood function, we develop a probabilistic estimation method that fully leverages the underlying measurement and detection models.
    Simulation results show that the proposed method significantly improves \gls{doa} estimation accuracy compared to baseline techniques, particularly in challenging scenarios with high missed detection rates. 
    Real-world experiments using \gls{ble} signals and directional antennas further validate the effectiveness of the approach, demonstrating substantial performance gains. 
    These findings highlight the value of modeling missed detections in sensor array processing and open new avenues for enhancing localization performance in wireless communication systems.
\end{abstract}

\end{minipage}}}

\maketitle

\glsresetall

\section{Introduction}
Accurate estimation of the \gls{doa} of a signal source is fundamental in many signal processing and communication systems, supporting tasks such as localization, tracking, and beamforming \cite{zetterqvistElephant2023,StillFUSION22AcousticLocalization,Werner2016RSSDoA}. The problem has been extensively studied \cite{Knapp:DoaStudy, Krim:DoaStudy, Stoica:DoaStudy}, with recent advances summarized in \cite{Rascon:RecentDoaStudy}.

Classical \gls{doa} methods rely on sensor arrays and exploit time delays (\gls{tdoa}) or spatial correlations, such as \gls{music} \cite{Schmidt:MUSIC} and \gls{esprit} \cite{Roy:ESPRIT}. More recent work improves robustness and performance in challenging conditions, including impulsive noise and limited array configurations \cite{Liu2023RobustDOAJSEN, Guo2024SparseDOA}.

In many practical systems, such as \gls{ble}, sensors provide only scalar \gls{rssi} measurements without phase coherence, making classical methods inapplicable. This has motivated \gls{rssi}- and power-based \gls{doa} approaches that exploit directional antenna or sensor gain patterns.

Several such methods have been proposed. For example, \cite{RssiDoaGmm2019} models \gls{rssi} measurements using Gaussian mixtures to localize sources from signal strength measurements. The Dir-MUSIC framework \cite{DirMusic2022, DirMusicWideBand2023} incorporates antenna gain patterns into a subspace formulation, enabling \gls{doa} estimation without phase synchronization; however, it still relies on multiple observations to form covariance estimates and requires both measurements and gain patterns in linear units rather than in dB. In acoustic sensing, \cite{zetterqvist2023, zetterqvistSensors2025} use directional sensitivity and received power measurements to estimate \gls{doa}, typically assuming continuous measurement availability. 

Machine learning approaches \cite{RssiLocalizationML2020, RssiLocalizationML20007, RssiLocalizationML2015} learn relationships from \gls{rssi} to location or direction using data-driven models, capturing complex propagation effects but requiring extensive training data and observed measurements during estimation.

% Despite these advances, existing methods generally assume that measurements are available and rely on aggregating observations. They do not explicitly model missed detections, where sensors fail to report measurements due to low signal strength, interference, or hardware limitations.

A common limitation of these methods is that they do not explicitly account for missed detections, where sensors fail to report measurements due to low signal strength, interference, or hardware limitations. In practice, this means that non-detections are either ignored or treated as missing data without information content.

Missed detections are common in wireless systems \cite{BLETrackingSurian2019, BLE_WiFi_Interference_Pang2022} and can provide useful information when detection thresholds are known. However, they are typically not exploited in existing \gls{doa} estimation approaches.

In this paper, we propose a probabilistic framework for \gls{rssi}-based \gls{doa} estimation that incorporates both detected signals and missed detections into the likelihood function. By jointly modeling detection probability and \gls{rssi} measurements, the method exploits information from both detections and non-detections.
The proposed method is evaluated through simulations and real-world \gls{ble} experiments with directional antennas, demonstrating improved performance in challenging conditions.

The main contributions of this paper are:
\begin{itemize}
    \item A probabilistic \gls{rssi}-based \gls{doa} estimation framework that explicitly incorporates the information encoded in missed detections.
    \item A likelihood formulation that jointly uses both detected signals and missed detections. 
    \item An evaluation using both simulated data and real-world \gls{ble} experiments with directional antennas.
\end{itemize}

This constitutes a modeling advancement for \gls{rssi}-based localization, improving estimation accuracy in scenarios with frequent missed detections caused by low signal strength, noise, or propagation effects.

% The method is evaluated through simulations and real-world experiments using \gls{ble} signals and directional antennas, demonstrating improved performance in challenging conditions.

The remainder of the paper is organized as follows: In \Secref{sec:problem_form}, the problem is formulated and the measurement model is presented.
This is followed by \Secref{sec:state_est_md}, where the proposed state estimation method that accounts for missed detections is described.
In \Secref{sec:application_doa}, the specific application of the proposed method to \gls{doa} estimation with directional sensors is derived.
In \Secref{sec:sim_study}, the simulation setup used to evaluate the proposed method is described, and the simulation results are presented.
\Secref{sec:exp_study} then presents the real-world experimental study, and the corresponding results are presented.
Finally, \Secref{sec:conclusions} concludes the paper and discusses future work.
\section{Problem Formulation}\label{sec:problem_form}
Throughout this paper, we consider a setting with $N$ sensor measurements, each originating from a sensing element characterized by a known measurement model $h_m(x)$, where~$x$ is the state to be estimated and $m = 1, 2, \ldots, N$ indexes the sensors. The index $m$ may correspond to different sensors in an array of directional elements or to multiple measurement instances from a single sensor. The sensors may share a common measurement model or each may have its own. 

The measurement model $h_m(x)$ is assumed to be known, and the specific form can depend on the sensing modality and sensor characteristics. No particular functional form is required, but $h_m(x)$ must provide variation with respect to the state $x$ in order to be informative for estimation. If the measurement model is almost flat, \ie, weakly dependent on $x$, it does not provide sufficient information for estimation and is therefore not suitable for the proposed method.

Each sensor measurement is associated with a known detection threshold $\gamma$. The objective is to estimate the state~$x$ of a signal source using the $n \le N$ measurements that result in detections. The state $x$ represents any parameter that influences the measurement model---such as \gls{doa}, position, or velocity---and is assumed to remain constant over the $N$ measurements. 

We further assume that a signal source is always present, so the state $x$ always exists, and that no false alarms occur; \ie, any detection made by a sensor is guaranteed to originate from the signal source whose state we aim to estimate.

Denote the sensor measurement, $Y$, then the measurement model is given by
\begin{equation}
   Y= \begin{cases}
        \smash{\overbrace{\{h(x) + e\}}^{\{y\}}}, & \text{if detected} \\
        \emptyset, & \text{otherwise,}
    \end{cases}
\end{equation}
where $h(x)$ is the measurement model of the state $x$, and $e$ is the measurement noise.
For simplicity, we assume that the measurement noise is normally distributed with zero mean and variance $\sigma^2$, \ie, $e \sim \mathcal{N}(0, \sigma^2)$.

\begin{figure}[tb]
    \centering
    \ifdefined\isCI
        \includegraphics{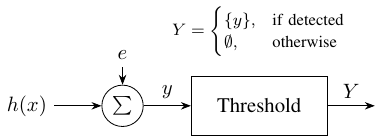}
    \else
        \tikzsetnextfilename{meas_model}
        \input{TiKz/meas_model.tex}
    \fi
    \caption{Illustration of the measurement model. 
      The measurement model, $h(x)$, and the noise, $e$, are added to the sum. 
      The output is then thresholded to determine if the signal was detected or not.}
    \label{fig:measurement_model}
\end{figure}
This measurement model can be illustrated as in \Figref{fig:measurement_model}, where the measurement model, $h(x)$, and the noise, $e$, are added to the sum. 
The output is then thresholded to determine if the signal was detected or not.
If the signal is detected, the output measurement is $Y = \{y\}$, otherwise the output is $Y = \emptyset$.
This is common in many wireless communication systems, where a receiver only processes signals that are above a certain threshold to reduce noise and interference.
For the case of \gls{ble} signals, the threshold $\gamma$ is typically set to $-95$~dBm.
\section{State Estimation with Missed Detections}\label{sec:state_est_md}
Given the above measurement model, we derive the likelihood function for the measurements from the sensors.
For the likelihood function, two cases need to be considered: when the signal is detected and when the signal is not detected.
In the calculations, we assume that the target exists, \ie, we do not consider the case of false alarms.

\begin{figure}[tb]
    \centering
    \ifdefined\isCI
        \includegraphics{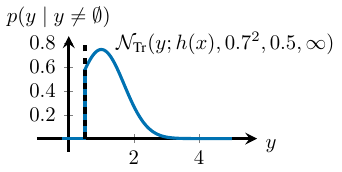}
    \else
        \tikzsetnextfilename{trunk_gauss}
        \input{TiKz/trunk_gauss.tex}
    \fi
    \caption{Illustration of the likelihood function for the case when the signal is detected, with $h(x) = 1$, $\sigma = 0.7$, and the threshold $\gamma = 0.5$. The dashed line indicates the threshold $\gamma$.}
  \label{fig:likelihood_detected}
\end{figure}

Denote the probability of detection as $p_D(x) = P(Y \neq \emptyset)$. The measurement model can then be expressed as:
\begin{equation}
    Y= \begin{cases}
        \{y\}, & \text{with probability } p_D(x) p(y|x) \\
        \emptyset, & \text{with probability } (1-p_D(x)) ,
    \end{cases}
\end{equation}
where $(1-p_D(x))$ is the probability of missed detection.
Breaking it down, we can express the likelihood function for the case when the signal is detected as a truncated normal distribution:
\begin{align}
    p(Y\mid x; Y\neq \emptyset) &= \mathcal{N}_{\text{Tr}}(Y; h(x), \sigma^2, \gamma, \infty),
\end{align}
where the truncated normal distribution is defined as:
\begin{equation}\label{eq:trunk_gauss}
    \mathcal{N}_{\text{Tr}}(Y; h(x), \sigma^2, a, b) = \frac{\mathcal{N}(Y; h(x), \sigma^2) \chi (a<Y<b)}{\int_a^b \mathcal{N}(u; h(x), \sigma^2) du},
\end{equation}
with $\chi(\cdot)$ being the indicator function.
A visualization of the likelihood function for the case when the signal is detected is shown in \Figref{fig:likelihood_detected}.

Thus, the total probability law can be expressed as:
\begin{align}
    p(Y\mid x) &= 
    \begin{cases}
        \mathcal{N}_{\text{Tr}}(Y; h(x), \sigma^2, \gamma, \infty) p_D(x), & \text{if detected} \\
        1-p_D(x), & \text{if not detected} .
    \end{cases}
\end{align}

Given the measurements from a set of sensors, denoting the set of sensors that detected the signal as $\mathcal{D}$ and the set of sensors that did not detect the signal as $\mathcal{MD}$, the resulting joint likelihood function is:
\begin{align}
    p(\mathbf{Y}\mid x) = & \prod_{m \in \mathcal{D}} \mathcal{N}_{\text{Tr}}(Y_m; h_m(x), \sigma^2, \gamma, \infty) p_{D,m}(x) \nonumber \\ 
    &\quad \cdot \prod_{m \in \mathcal{MD}} (1-p_{D,m}(x)) ,
\end{align}
where $\mathbf{Y}$ is the set of measurements from all sensors, $m$ is the sensor index, $h_m(x)$ is the measurement model for sensor~$m$, and $p_{D,m}(x)$ is the probability of detection for sensor $m$, given measurement model $h_m(x)$.
\begin{figure}[tb]
    \centering
    \ifdefined\isCI
        \includegraphics{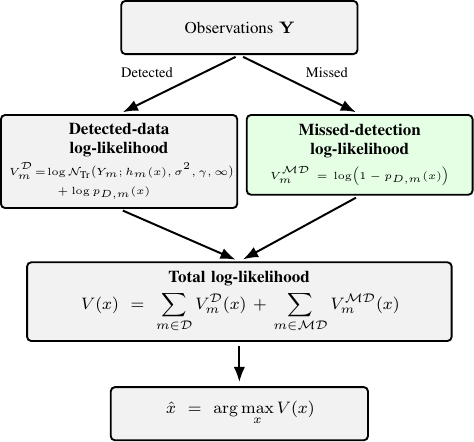}
    \else
        \tikzsetnextfilename{flowchart}
    \input{TiKz/flowchart.tex}
    \fi
    \caption{Overview of the proposed estimation method. The observations are partitioned into detected and missed measurements. These contribute separate detected-data and missed-detection log-likelihood terms, which are combined into a total log-likelihood. The final parameter estimates are obtained by maximizing the objective over $x$.
    The missed-detection log-likelihood term is the key novelty of the proposed method, as it allows for a more accurate estimation of the state by leveraging the information from missed detections.}
    \label{fig:flowchart}
\end{figure}

When multiple observations are available over time, assuming conditional independence between snapshots, the likelihood can be extended as:
\begin{align}
    p(\mathbf{Y}\mid x) = \prod_{t=1}^T p(\mathbf{Y}_t\mid x),
\end{align}
where $\mathbf{Y}_t$ denotes the set of measurements from all sensors at time $t$, and $T$ is the number of snapshots.

For the purpose of estimating the state $x$, we can use the \gls{ml} estimate of the parameters $x$.
This is done by maximizing the likelihood function with respect to the parameters, which is expressed as:
\begin{align}
    \hat{x} &= \arg\max_{x} p(\mathbf{Y}\mid x) .
\end{align}
For ease of optimization, this is equivalently formulated as the minimization of the negative log-likelihood function:
\begin{alignat}{2}
\hat{x} 
    &= \arg\min_x && \bigl[-\log p(\mathbf{Y}\mid x)\bigr] \\[1ex]
    &= \arg\min_x && \Biggl\{ 
        - \sum_{m \in \mathcal{D}} 
            \Bigl[
                \log \mathcal{N}_{\text{Tr}}\bigl(
                    Y_m;
                    h_m(x),\,
                    \sigma^2,\,
                    \gamma,\,
                    \infty
                \bigr)  \nonumber 
                \\ & && \qquad + \log p_{D,m}(x)
            \Bigr] \nonumber  \\ \label{eq:loglikelihood_x}
    &&&\quad - \sum_{m \in \mathcal{MD}} 
            \log\!\left( 1 - p_{D,m}(x) \right)
    \Biggr\}.
\end{alignat}
This cost function can then be minimized using standard optimization techniques, such as gradient descent or Newton's method, to obtain the \gls{ml} estimate of the state $x$.
For dynamic scenarios, the same likelihood formulation can be incorporated into a Bayesian filtering framework, \eg, a \gls{pf}, for recursive state estimation.

Note that in this formulation, the \gls{fim} and the resulting \gls{crlb} for the estimation problem cannot be derived in closed form, due to the non-differentiable nature of the likelihood function with respect to the state $x$ caused by the thresholding operation in the measurement model.

A flowchart illustrating the proposed estimation method is shown in \Figref{fig:flowchart}. The novelty of the proposed method lies in the explicit modeling of missed detections in the likelihood function, which allows for a more accurate estimation of the state $x$ by leveraging both detected and missed measurements from the sensors.
\section{Application to DOA Estimation}\label{sec:application_doa}
The proposed state estimation method that accounts for missed detections can be applied to various estimation problems, such as localization, tracking, and parameter estimation.
In this section, we focus on the application of the proposed method to \gls{doa} estimation using directional sensors, which is a common problem in wireless localization and tracking systems.

One such application is \gls{rssi}-based \gls{doa} estimation using directional antennas, where the sensors measure the received signal strength from a signal source at an unknown direction.
In this context, the state to be estimated is the \gls{doa} angle $\psi$ of the signal source, and the measurement model is given by 
\begin{equation}\label{eq:meas_model_sim}
    y_m = \underbrace{\alpha + h_m(\psi)}_{=h_m(x)} + e_m, \quad m = 1, 2, \ldots, N,
\end{equation}
where $\alpha$ denotes the signal power at the center of the sensor array, assumed identical for all sensors, $h_m(\psi)$ represents the directional sensitivity pattern of sensor $m$ as a function of the \gls{doa} angle $\psi$, and $e_m$ is the measurement noise at sensor $m$. 
The noise is assumed to follow a Gaussian distribution with zero mean and variance $\sigma^2$, \ie, $e_m \sim \mathcal{N}(0, \sigma^2)$. 

The directional sensitivity pattern of each sensor is modeled using a \gls{fs} with $K$ harmonics, given by
\begin{equation}\label{eq:fs_model}
    h_m(\psi) = \sum_{k=-K}^{K} c_{m,k} \, e^{i k \psi},
\end{equation}
where $c_{m,k}$ are the complex Fourier coefficients for sensor $m$.
A \gls{fs} is a natural choice for modeling the directional sensitivity pattern of the sensors, as it can capture the periodic nature of the response and can be easily parameterized using a finite number of harmonics.

With this setup, the optimization problem from \eqref{eq:loglikelihood_x} in \Secref{sec:state_est_md} can be reformulated to jointly estimate the \gls{doa} angle $\psi$ and the signal power $\alpha$ as
\begin{align}
    \hat{\psi}, \hat{\alpha} &= \arg\min_{\psi, \alpha} \Biggl\{\sum_{m \in \mathcal{D}}
    \Biggl[ 
    \frac{(Y_m - (\alpha + h_m(\psi)))^2}{2\sigma^2}  \nonumber \\
    &\qquad\qquad + \log \left(1 - \Phi\left(\frac{\gamma - (\alpha + h_m(\psi))}{\sigma}\right)\right) \nonumber \\
    &\qquad\qquad - \log(p_{D,m}(\psi, \alpha)) \Biggr] \nonumber \\
    & \qquad - \sum_{m \in \mathcal{MD}} \log \left(1 - p_{D,m}(\psi, \alpha)\right) \Biggr\} , \label{eq:sim_cost_function}
\end{align}
where we have inserted the definition of the truncated normal distribution from \eqref{eq:trunk_gauss}, and $\Phi(\cdot)$ denotes the \gls{cdf} of the normal distribution.

The probability of detection is conveniently modeled as
\begin{align}
    p_{D,m}(\psi, \alpha) = p_{c,m} \cdot p_{\alpha,m}(\psi, \alpha) , \label{eq:prob_detect_total}
\end{align}
where $p_{c,m}$ is a constant detection efficiency term corresponding to the standard probability of detection used in conventional tracking frameworks, \ie, the probability of detecting a target-generated signal. The term $p_{\alpha,m}(\psi, \alpha)$ represents the detection probability due to thresholding.

The threshold-based detection probability is determined by the detection threshold $\gamma$ and given by
\begin{align}
    p_{\alpha,m}(\psi, \alpha) = 1 - \Phi\left(\frac{\gamma - (\alpha + h_m(\psi))}{\sigma}\right) . \label{eq:prob_detect}
\end{align}

This factorization separates ideal threshold-based detection from additional loss mechanisms and is consistent with standard probability-of-detection models used in tracking frameworks. The resulting optimization problem is then given by
\begin{align}
\hat{\psi}, \hat{\alpha} 
&= \arg\min_{\psi, \alpha}
\Bigg\{
    \sum_{m \in \mathcal{D}} \left[ \frac{(Y_m - (\alpha + h_m(\psi)))^2}{2\sigma^2} - \log(p_{c,m}) \right] \nonumber\\
&\qquad
    - \sum_{m \in \mathcal{MD}} 
        \log\!\left(
            1 - p_{c,m} \Phi\left(\frac{(\alpha + h_m(\psi)) - \gamma}{\sigma}\right)
        \right)
\Bigg\}, \label{eq:sim_cost_function_final}
\end{align}
where the first term corresponds to the contribution from the detected signals, and the second term corresponds to the contribution from the missed detections.

\section{Simulation Study}\label{sec:sim_study}
In this section, the performance of the proposed state estimation method that accounts for missed detections is evaluated through simulations.

\subsection{Simulation Setup}
For the experiments, we consider a scenario with $N=4$ sensors arranged in a \gls{uca} configuration. 
The setup is strongly motivated by our real-world \gls{ble} antenna array, which is presented later in \Secref{sec:exp_study}. 

The sensors measure the signal strength from a single source whose state (the \gls{doa} angle $\psi$ and the signal power $\alpha$) is unknown, leading to the measurement model presented in \eqref{eq:meas_model_sim}. The directional sensitivity pattern of each sensor is modeled using a \gls{fs} with $K=7$ harmonics, as given in \eqref{eq:fs_model}.

\begin{figure}[tb]
    \centering
    \ifdefined\isCI
        \includegraphics{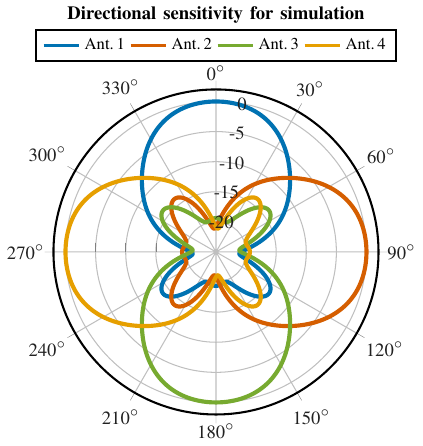}
    \else
        \tikzsetnextfilename{antenna_patterns_simulated}
        \input{TiKz/antenna_patterns_simulated.tex}
    \fi
    \caption{Simulated sensor array configuration and the directional sensitivity patterns for each sensor. 
    The sensors are placed in a \gls{uca} configuration, and each sensor has a unique directional sensitivity pattern modeled as a \gls{fs}.}
    \label{fig:antenna_patterns_simulated}
\end{figure}

The sensor array configuration and the directional sensitivity patterns are illustrated in \Figref{fig:antenna_patterns_simulated}.
The patterns are modeled after real-world sensor 
measurements presented in \Secref{sec:exp_study}, featuring a main lobe in 
the boresight direction, side lobes, and a back lobe to capture the sensor's full angular response.

The optimization is performed using a grid search over the possible values of $\psi$ and $\alpha$, with $\psi$ ranging from $0^\circ$ to $360^\circ$ in $1^\circ$ increments, and $\alpha$ ranging from $-100$\,dBm to $0$\,dBm in $0.2$\,dBm increments.

For experimental comparison, we consider a baseline estimator based on the standard \gls{nls} formulation commonly used in state and parameter estimation problems \cite{kay:estimation}. In this approach, the \gls{doa} is estimated using only the available \gls{rssi} measurements from successful detections, while missed detections are ignored. Similar formulations are commonly used in \gls{rssi}-based \gls{doa} estimation using directional antennas \cite{NLSDirSensDoA2018}. This baseline therefore represents the conventional approach where only detected measurements are used in the estimation process.

\subsection{Simulation Results}\label{sec:sim_results}
To evaluate the performance of the proposed state estimation method that accounts for missed detections, we conduct a series of simulations.
The simulation setup follows the method described in \Secref{sec:application_doa}.
The true \gls{doa} angle $\psi$ is varied over the range $[-180^\circ, 180^\circ]$, and different levels of $\alpha$ are considered to simulate different signal strengths, and thus different probabilities of detection.
To focus on the impact of the threshold-based detection and missed detections, the constant detection efficiency term $p_{c,m}$ is set to $1$ for all sensors, representing an ideal scenario where there are no additional loss mechanisms beyond the threshold-based detection.
The threshold $\gamma$ is set to $-95$\,dBm, and the noise
variance $\sigma^2$ is set to $4$\,dBm$^2$.
For each configuration of $\psi$ and $\alpha$, we generate $50$ \gls{mc} runs to evaluate the performance of the proposed method.
\begin{figure}[tb]
    \centering
    \ifdefined\isCI
        \includegraphics{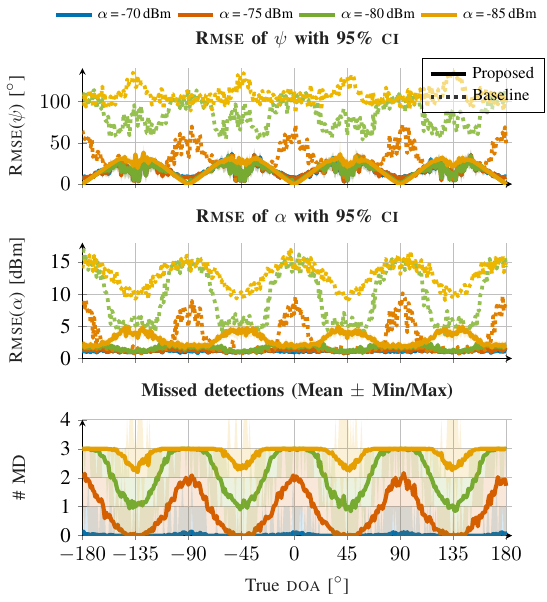}
    \else
        \tikzsetnextfilename{sim_rmse_alpha_results_vs_base}
        \input{TiKz/sim_rmse_alpha_results_vs_base.tex}
    \fi
    \caption{Simulation results showing the \gls{rmse} of the \gls{doa} estimates as a function of the true \gls{doa} angle $\psi$ for different levels of $\alpha$ in the top plot. 
    The mid plot shows the \gls{rmse} of the $\alpha$ estimates, and the bottom plot shows the average number of missed detections.
    The shaded areas in the top and mid plots represent the 95\% \gls{ci} of the estimates over the 50 \gls{mc} runs, while the bottom plot area shows the minimum and maximum number of missed detections over the \gls{mc} runs.}
    \label{fig:rmse_results}
\end{figure}

\begin{figure}[tb]
    \centering
    \ifdefined\isCI
        \includegraphics{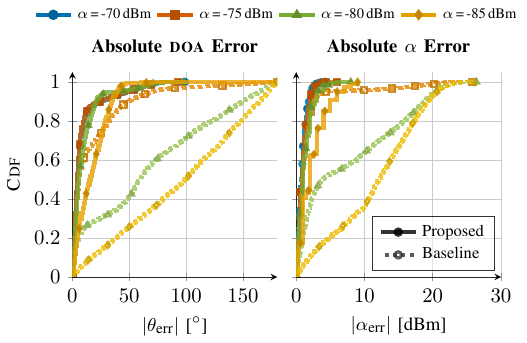}
    \else
        \tikzsetnextfilename{CDF_doa_and_alpha}
        \input{TiKz/CDF_doa_and_alpha.tex}
    \fi
    \caption{\Glsxtrfull{cdf} of the \gls{doa} estimation error in the left plot and $\alpha$ estimation error in the right plot for different levels of $\alpha$. The proposed method is shown with solid lines, while the baseline method is shown with dashed lines.}
    \label{fig:cdf_results}
\end{figure}

The performance is evaluated in terms of the \gls{rmse} of the \gls{doa} and $\alpha$ estimates, and compared against the baseline method that only uses the detected signals for \gls{doa} estimation.
The results of the simulations are presented in \Figref{fig:rmse_results}, which shows the \gls{rmse} of the \gls{doa} estimates as a function of the true \gls{doa} angle $\psi$ for different levels of $\alpha$, as well as the \gls{rmse} of the~$\alpha$ estimates and the average number of missed detections.
In \Tabref{tab:rmse_summary}, the mean \gls{rmse} of the \gls{doa} estimates across all angles for different levels of $\alpha$ is summarized.
\begin{table}[tb]
    \centering
    \caption{Summary of the mean \gls{rmse} of the \gls{doa} and $\alpha$ estimates across all angles for different levels of $\alpha$, with $\pm$1 standard deviation shown as subscripts. Bold indicates the better performance for each $\alpha$ level.}
    \begin{tabular}{c c c c c}
    \toprule
    \multirow{2}{*}{$\alpha$ [dBm]} & \multicolumn{2}{c}{Mean \glsxtrshort{doa} \glsxtrshort{rmse} [\textdegree]} & \multicolumn{2}{c}{Mean $\alpha$ \glsxtrshort{rmse} [dBm]} \\
    \cmidrule(r){2-3} \cmidrule(r){4-5}
    & Proposed & Baseline & Proposed & Baseline \\
    \midrule
    \phantom{$^\star$}-70$^\star$ & \textbf{16.2} \tiny $\pm$ 2.8 & 16.5 \tiny $\pm$ 2.8 & 1.13 \tiny $\pm$ 0.16 & \textbf{1.12} \tiny $\pm$ 0.16 \\
    -75 & \textbf{16.0} \tiny $\pm$ 2.7 & 31.0 \tiny $\pm$ 5.3 & \textbf{1.24} \tiny $\pm$ 0.17 & 2.93 \tiny $\pm$ 0.37 \\
    -80 & \textbf{15.3} \tiny $\pm$ 2.2 & 84.5 \tiny $\pm$ 6.9 & \textbf{1.63} \tiny $\pm$ 0.22 & 9.58 \tiny $\pm$ 0.84 \\
    -85 & \textbf{19.3} \tiny $\pm$ 0.8 & 107.8 \tiny $\pm$ 6.4 & \textbf{2.96} \tiny $\pm$ 0.35 & 12.90 \tiny $\pm$ 0.96 \\
    \bottomrule
    \multicolumn{5}{l}{\tiny $^\star$ Note: For $\alpha = -70$ dBm, the difference is not statistically significant.}\\
\end{tabular}

    \vspace{-0.5cm}
    \label{tab:rmse_summary}
\end{table}
The \gls{cdf} of the \gls{doa} estimation error and the $\alpha$ estimation error is also computed for different values of $\alpha$, as shown in \Figref{fig:cdf_results}.
From the results, it is observed that for high values of $\alpha$ (\eg, $-70$\,dBm), both the proposed method and the baseline method perform similarly, as the probability of detection is high, resulting in few missed detections.
However, as $\alpha$ decreases, the proposed method significantly outperforms the baseline method, demonstrating its effectiveness in handling missed detections.

\begin{figure}[tb]
    \centering
    \ifdefined\isCI
        \includegraphics{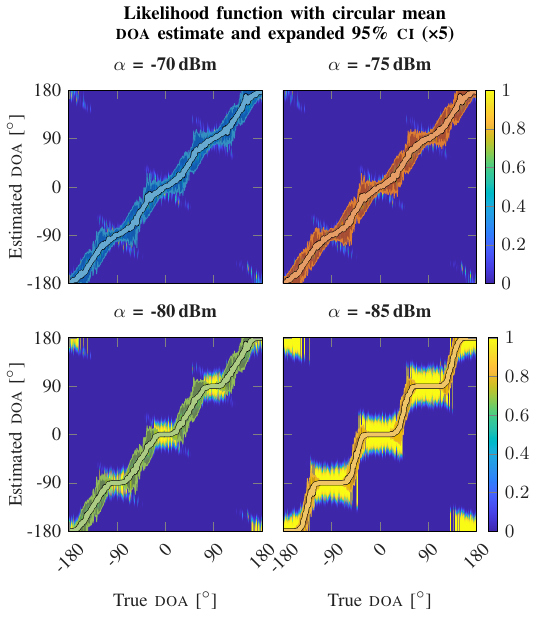}
    \else
        \tikzsetnextfilename{cost_fcn_alpha_CI}
        \input{TiKz/cost_fcn_alpha_CI.tex}
    \fi
    \caption{The cost function for \gls{doa} estimation as a function of the angle $\psi$ for different levels of $\alpha$. The mean \gls{doa} estimate from the proposed method is indicated by the solid line, and the shaded area represents the 95\% \gls{ci} (expanded by a factor of 5 for visibility) of the estimates over the 50 \gls{mc} runs. The cost function is from a single \gls{mc} run, showing the multiple minima around certain angles due to the symmetry of the directional sensitivity patterns of the sensors.}    
    \label{fig:cost_function}
\end{figure}

It is also observed that the performance of the estimator degrades at certain angles, more specifically around $\psi = \pm 45^\circ$ and $\psi = \pm 135^\circ$.
This is due to the directional sensitivity patterns of the sensors, which are symmetric, leading to ambiguities in the \gls{doa} estimation.
This is also reflected in the resulting cost function, which exhibits multiple minima around these angles, making it challenging for the optimization algorithm to converge to the correct solution, as illustrated in \Figref{fig:cost_function}.
Overall, the simulation results validate the effectiveness of the proposed state estimation method that accounts for missed detections, demonstrating its potential for improving \gls{doa} estimation accuracy in practical scenarios with sensor arrays.

\begin{figure}[tb]
    \centering
    \ifdefined\isCI
        \includegraphics{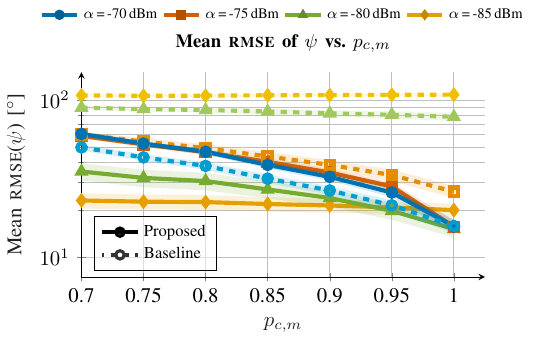}
    \else
        \tikzsetnextfilename{sim_rmse_vs_pcm}
        \input{TiKz/sim_rmse_vs_pcm.tex}
    \fi
    \caption{Simulation results showing the mean \gls{rmse} of the \gls{doa} estimates as a function of the constant detection efficiency term $p_{c,m}$ for different levels of $\alpha$. The shaded areas represent the 95\% \gls{ci} of the estimates over the 50 \gls{mc} runs.
    }
    \label{fig:rmse_vs_pcm}
\end{figure}

In practical applications, the constant detection efficiency term $p_{c,m}$ is typically less than $1$ due to hardware imperfections, environmental conditions, and other loss mechanisms. To evaluate the robustness of the proposed method under such conditions, simulations are conducted for $p_{c,m} \in [0.7, 1]$, assuming identical values across all sensors. 
The results are shown in \Figref{fig:rmse_vs_pcm}, which presents the mean \gls{rmse} of the \gls{doa} estimates as a function of $p_{c,m}$ for different levels of $\alpha$. 
The corresponding $\alpha$ estimates and the average number of missed detections for varying $p_{c,m}$ are also evaluated, but omitted here for brevity.
Note that the rightmost data points in \Figref{fig:rmse_vs_pcm} correspond to the case with $p_{c,m}=1$, matching the results in \Tabref{tab:rmse_summary}.

The results show that the proposed method remains robust to reduced detection efficiency, outperforming the baseline for lower values of $\alpha$. However, the performance degrades for higher values of $\alpha$ compared to the case with $p_{c,m} = 1$, which is expected since reduced detection efficiency leads to more missed detections even at higher signal strengths.

Notably, for $\alpha=-70$\,dBm and $p_{c,m}<1$, the proposed method performs worse than the baseline. This occurs because~$p_{c,m}<1$ introduces missed detections even within the nominal detection range. 
Since the proposed method exploits missed detections as informative events, these additional, non-threshold-dependent missed detections introduce bias in the estimates.
For small sample sizes or few sensor elements, this bias can result in worse performance compared to the baseline method, which does not explicitly model missed detections and is therefore more robust to such biases in this specific scenario. Nevertheless, for lower values of $\alpha$, the proposed method continues to outperform the baseline method, demonstrating robustness in scenarios with more frequent missed detections.

\begin{figure}[tb]
    \centering
    \ifdefined\isCI
        \includegraphics{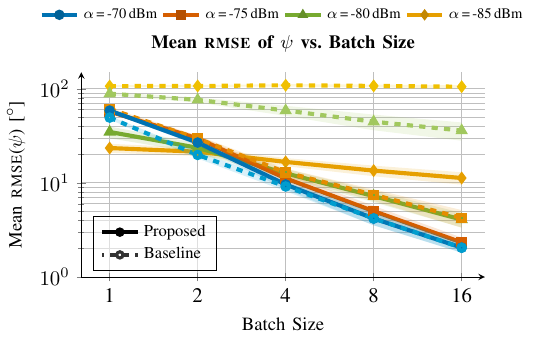}
    \else
        \tikzsetnextfilename{sim_rmse_vs_BatchSize}
        \input{TiKz/sim_rmse_vs_BatchSize.tex}
    \fi
    \caption{Simulation results showing the mean \gls{rmse} of the \gls{doa} estimates as a function of the batch size for different levels of $\alpha$ using $p_{c,m}=0.7$. The shaded areas represent the 95\% \gls{ci} of the estimates over the 50 \gls{mc} runs.}
    \label{fig:rmse_vs_BatchSize}
\end{figure}

To examine how this effect evolves with increasing data, simulations are also conducted with varying batch sizes, ranging from $1$ to $16$ measurements per sensor, while keeping~$p_{c,m}=0.7$ constant.
The results are shown in \Figref{fig:rmse_vs_BatchSize}, which presents the mean \gls{rmse} of the \gls{doa} estimates as a function of the batch size for different levels of $\alpha$.
Note that the leftmost data points are the same in \Figref{fig:rmse_vs_BatchSize} and \Figref{fig:rmse_vs_pcm}, as they correspond to the same setup with a batch size of $1$, and $p_{c,m}=0.7$.

The results indicate that increasing the number of measurements mitigates the bias introduced by detection inefficiencies. Consequently, the proposed method outperforms the baseline for lower values of $\alpha$, while achieving comparable performance for higher values of $\alpha$, where missed detections are less frequent and less informative. This demonstrates that the method can effectively leverage additional information, including missed detections, when enough data are available.

In scenarios where the number of measurements per sensor is limited, \eg, due to time constraints or a moving source, similar performance can be achieved by increasing the number of sensors. This effectively increases the total amount of information available for estimation, helping to offset the bias introduced by non-ideal detection efficiency.

\section{Experimental Study}\label{sec:exp_study}
In this section, we present the experimental setup for validating the proposed state estimation method that accounts for missed detections using real-world \gls{ble} signals and directional antennas.

\subsection{Experimental Setup}\label{sec:exp_design}
\begin{figure}[tb]
    \centering
    \includegraphics[width=0.49\columnwidth]{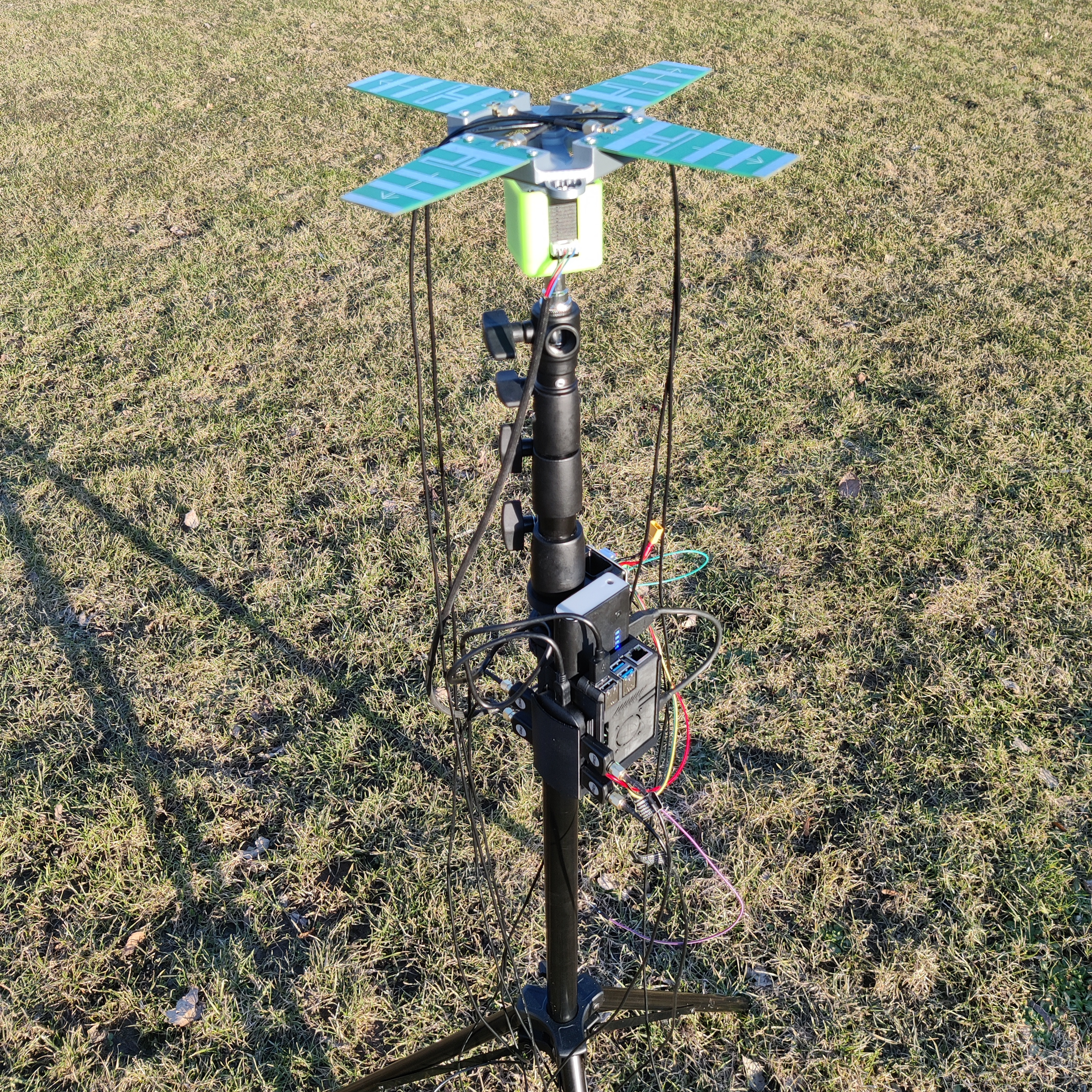}
    \includegraphics[width=0.49\columnwidth]{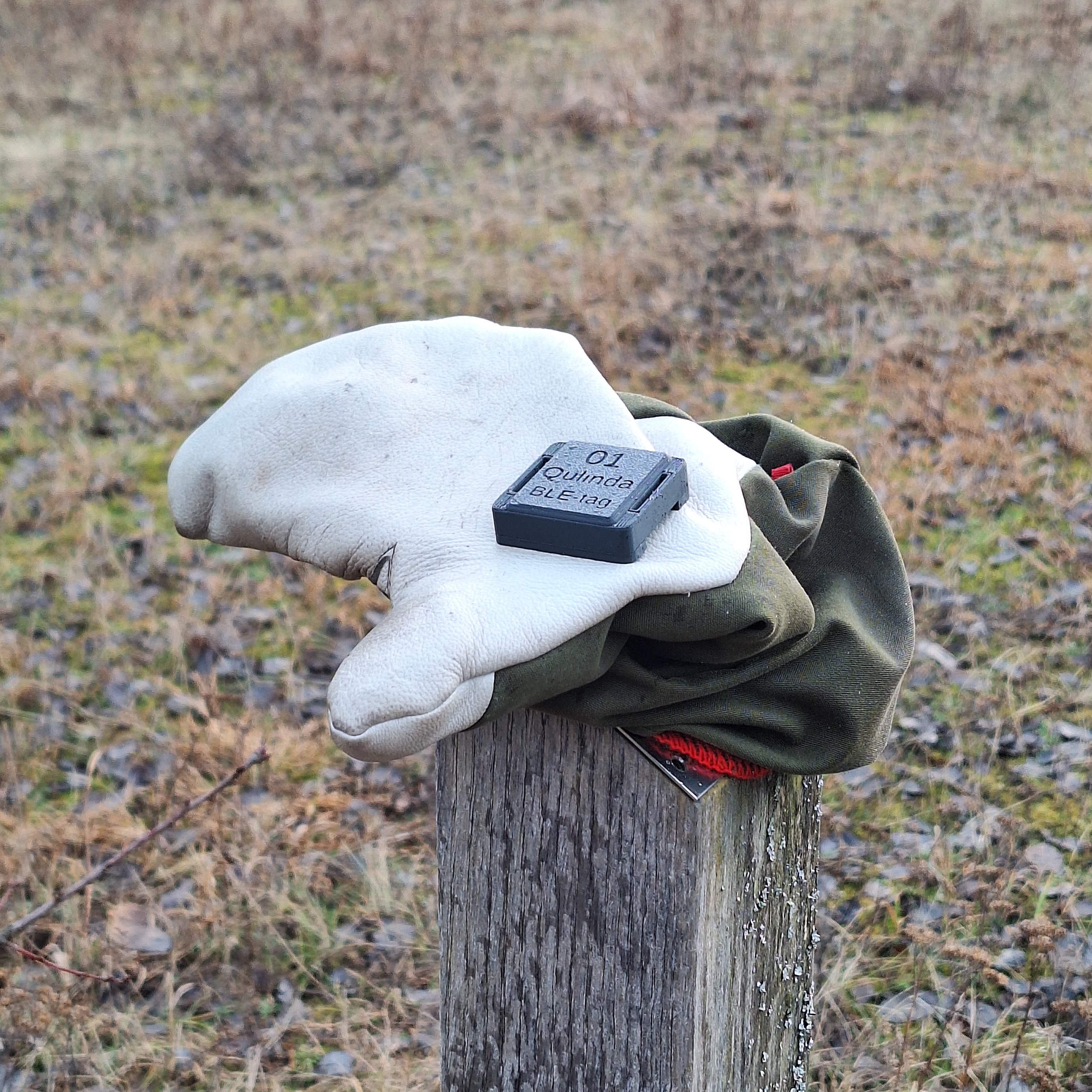}
    \caption{Pictures of the \gls{ble} Yagi antenna array used in the experiments (left) and the \gls{ble} beacon (right). 
    The array consists of four directionally sensitive \gls{ble} Yagi antennas arranged in a \gls{uca} configuration, connected to a Raspberry Pi 5 for data collection.}
    \label{fig:ble_array}
\end{figure}

For the real-world experiments, we utilize a setup consisting of $N=4$ directionally sensitive \gls{ble} Yagi antennas arranged in a \gls{uca} configuration, resulting in a difference in mounting angle of $90^\circ$ between each antenna. 
The antennas are connected to a Raspberry Pi 5, which is used to collect the \gls{rssi} measurements from a \gls{ble} beacon via a nRF52840 USB \gls{ble} dongle. 
The array is mounted on a rotating stepper motor, allowing us to vary the \gls{doa} angle $\psi$ of the incoming signal with respect to the sensor array. 
The \gls{ble} beacon used in the experiments is a Qulinda \gls{ble} tag transmitting \gls{ble} advertising packets.
A picture of the \gls{ble} Yagi antenna array and the \gls{ble} beacon is shown in \Figref{fig:ble_array}. 

For the \gls{ble} signal, data packets are transmitted at a rate of $10$\,Hz from the \gls{ble} beacon, using the three advertising channels (37, 38, and 39). During the experiments, it was observed that the directional sensitivity patterns of the antennas varied slightly between the different advertising channels. 
To account for this, we measure the directional sensitivity patterns for each antenna and advertising channel combination, and view the measurements as independent sensors. 
This results in a total of $N=12$ sensors (4 antennas $\times$ 3 advertising channels) in the sensor array for the experiments.

To characterize the directional sensitivity patterns of the antennas, we perform measurements in an outdoor environment with minimal multipath effects. 
The sensor array is rotated in $1^\circ$ increments over a full $360^\circ$ rotation, and the \gls{rssi} measurements are recorded at each angle for all antennas and advertising channels, with the \gls{ble} beacon placed at a fixed distance of $1$~meter from the sensor array. 
To obtain the directional sensitivity patterns, a minimum of $100$ \gls{rssi} measurements are collected at each angle for each antenna across three advertising channels. For each angle-antenna-channel combination, the mean and variance of the \gls{rssi} measurements are estimated. A \gls{fs} model is subsequently fitted to the mean values using \gls{wls}, where the weights are determined by the measurement variance.

\begin{figure}[tb]
    \centering
    \ifdefined\isCI
        \includegraphics{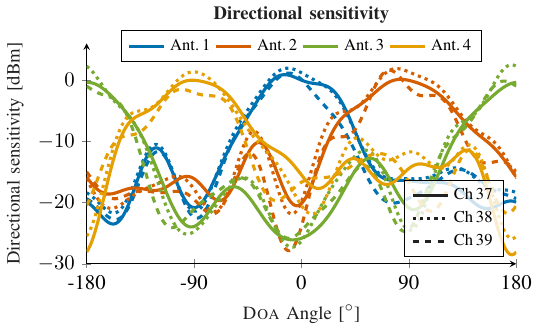}
    \else  
        \tikzsetnextfilename{DirSensReal}
        \input{TiKz/DirSensReal.tex}
    \fi
    \caption{\Gls{fs} models of the directional sensitivity patterns of the four \gls{ble} Yagi antennas used in the real-world experiments using an order of $K=7$ harmonics. 
    The patterns exhibit a main lobe in the boresight direction and side lobes, similar to the patterns used in the simulations.}
    \label{fig:antenna_patterns}
\end{figure}

The \gls{fs} models of the directional sensitivity patterns for the four antennas are shown in \Figref{fig:antenna_patterns}, using an order of~$K=7$ harmonics. 
Here, $K$ corresponds to the order of the \gls{fs} expansion and determines the level of detail captured in the directional pattern. 
The choice of $K=7$ provides a good balance between model complexity and estimation accuracy, based on empirical observations. 
Higher values of $K$ can represent more complex patterns but increase computational cost and the risk of overfitting. 
In our previous work \cite{Sundvall_Olsson_2025}, a \gls{bic} analysis was used to select~$K$, and $K=7$ was found to be a suitable choice for our data.

The patterns exhibit a main lobe in the boresight direction and side lobes, similar to the simulated patterns used in the simulations.
These measured patterns are then used as the measurement models, $h_m(\psi)$, for each sensor in the state estimation process. 
More details about the training of the \gls{fs} model can be found in \cite{zetterqvistSensors2025,Sundvall_Olsson_2025}.

\begin{figure}[tb]
    \centering
    \ifdefined\isCI
        \includegraphics{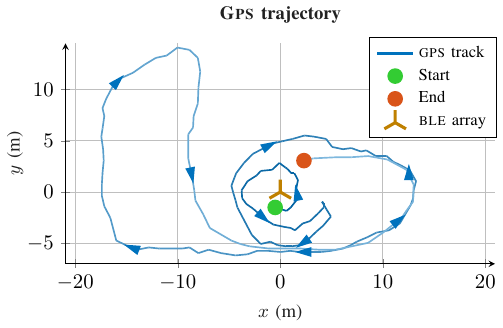}
    \else
        \tikzsetnextfilename{GPSTrack}
        \input{TiKz/GPSTrack.tex}
    \fi
    \caption{\Gls{gps} trajectory used in the real \gls{ble} data experiments. 
    The track gradually shifts to lighter blue as time progresses.}
    \label{fig:gps_track}
\end{figure}

For validation of the proposed state estimation method, we perform experiments where a person carrying the \gls{ble} beacon walks around the sensor array at varying distances and angles.
The true distance and \gls{doa} angle $\psi$ were measured using the \gls{gps} on an iPhone 15 Pro. 
The experiments were conducted in an open outdoor area to minimize multipath effects, and the sensor array is placed at a fixed location while the person carrying the \gls{ble} beacon walks around it.
An illustration of the \gls{gps} trajectory from one of the experiments is shown in \Figref{fig:gps_track}.

To evaluate the performance of the proposed method under conditions with frequent missed detections, we introduced a higher effective detection threshold of $\gamma = -65$\,dBm during the analysis stage. While the \gls{ble} receiver sensitivity is $-95$\,dBm, increasing the effective threshold allows us to emulate scenarios with reduced detection capability and systematically study the impact of missed detections on estimation performance.
\subsection{Experimental Results}\label{sec:exp_results}
Using the real \gls{ble} data with missed detections, we can further evaluate the performance of the proposed method. 
By applying the same optimization framework to the real data, we can assess the impact of missed detections on the \gls{doa} estimates and the overall system performance.

\subsubsection{Real BLE Data and Challenges}
When working with real data, it is important to consider environmental factors that affect the \gls{rssi} measurements, such as multipath propagation, interference from other devices, and obstacles in the signal path. 
These factors introduce additional variability into the measurements, making accurate \gls{doa} estimation more challenging.

One such effect is the occurrence of false negatives, where a sensor fails to detect a signal that is actually present. 
This can arise for various reasons, such as low signal strength, interference, or hardware limitations. 
Within the proposed framework, this is captured by the detection efficiency $p_{c,m}$, which models additional losses due to practical effects such as interference, hardware limitations, or unmodeled propagation effects. 
The resulting probability of detection is given by
\begin{equation}
    p_{D,m}(x) = p_{c,m}\, p_{\alpha,m}(x),
\end{equation}
where $p_{\alpha,m}(x)$ denotes the nominal detection probability due to thresholding. The resulting optimization problem remains the same as in \eqref{eq:sim_cost_function_final}.
\begin{figure}[tb]
    \centering
    \ifdefined\isCI
        \includegraphics{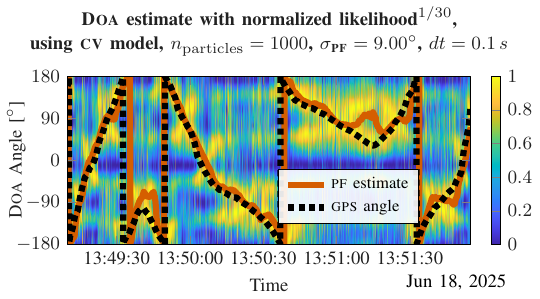}
    \else
        \tikzsetnextfilename{DoaPFEsti_X30_Sigma9_t01_w_likelihood_0}
        \input{TiKz/DoaPFEsti_X30_Sigma9_t01_w_likelihood_0.tex}
    \fi
    \caption{The resulting \gls{doa} estimates from real \gls{ble} data with missed detections overlayed with the likelihood function. A \gls{pf} is used to track the correct peak in the likelihood function. The ground truth \gls{doa} measurements from the \gls{gps} data are also shown for comparison.}
    \label{fig:RealDataResults}
\end{figure}
\begin{figure*}[tb]
    \centering
    \subfloat[$\gamma = -95$\,dBm]{
        \ifdefined\isCI
            \includegraphics{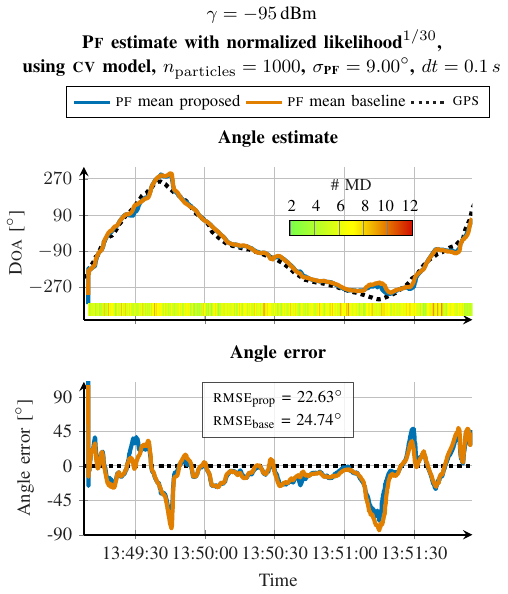}
        \else
            \tikzsetnextfilename{DoaPFEsti_X30_Sigma9_t01_w_errors_w_base_T95-only-angle}
            \input{TiKz/DoaPFEsti_X30_Sigma9_t01_w_errors_w_base_T95-only-angle.tex}
        \fi
        \label{fig:RealDataError}
    }
    % \hfill
    \subfloat[$\gamma = -65$\,dBm]{
        \ifdefined\isCI
            \includegraphics{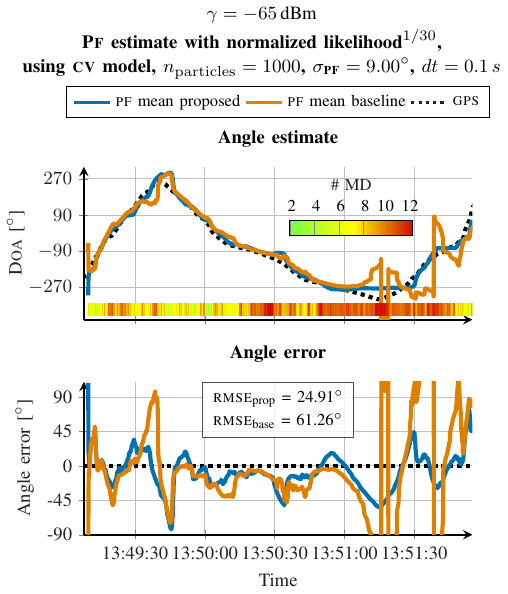}
        \else
            \tikzsetnextfilename{DoaPFEsti_X30_Sigma9_t01_w_errors_w_base_T65-only-angle}
            \input{TiKz/DoaPFEsti_X30_Sigma9_t01_w_errors_w_base_T65-only-angle.tex}
        \fi
        \label{fig:HighMissedDetRealDataResults}
    }
    \caption{The resulting estimation error from real \gls{ble} data with missed detections for the \gls{doa}, using both the proposed method and the baseline method. 
    A \gls{pf} is used to track the correct peak in the likelihood function. 
    The ground truth \gls{doa} measurements from the \gls{gps} data are also shown for comparison, as well as the number of missed detections over time in the colorbar in the upper plots. In (\protect\subref*{fig:RealDataError}) the nominal detection threshold of $\gamma = -95$\,dBm, and in (\protect\subref*{fig:HighMissedDetRealDataResults}) an increased detection threshold of $\gamma = -65$\,dBm.
    }\label{fig:realPF}
\end{figure*}
Unlike the simulations in \Secref{sec:sim_study}, where a fixed value of $p_{c,m}$ was used for all sensors, the detection efficiency in the real data is estimated from observed missed detection rates and may vary across sensors, channels, and time. In the real \gls{ble} experiments, we observed such variations in false negative probabilities, with channel~37 exhibiting a higher rate than the other channels. These estimates of $p_{c,m}$ are then used to compensate for missed detections in the proposed method.

The measurement noise variance $\sigma^2$ is also estimated from the real data by examining the variability of the \gls{rssi} measurements when the source is known to be present.

\subsubsection{DOA Estimation with Particle Filtering}
By incorporating this compensation for false negatives into the proposed method, we can improve the robustness of the \gls{doa} estimates in the presence of missed detections. 
The resulting \gls{doa} estimates can be compared to ground truth measurements obtained from \gls{gps} data, allowing us to evaluate the accuracy and reliability of the proposed method in real-world scenarios.

Furthermore, it was observed that simply taking the maximum of the likelihood function for \gls{doa} estimation could lead to incorrect peak selection due to the presence of multiple peaks in the cost function, especially in scenarios with high missed detection rates. 
To address this issue, we implemented a \gls{cv} \glsxtrfull{pf} that utilizes the likelihood function as measurements. 
This approach allows us to track the correct peak over time, improving the accuracy of the \gls{doa} estimates.

The results of applying the proposed method with false negative compensation and the \gls{pf} to the real \gls{ble} data are presented in \Figref{fig:RealDataResults}. 
The figure shows the \gls{doa} estimates from the \gls{pf}. 
The ground truth \gls{doa} measurements from the \gls{gps} data are also included for comparison, as well as the likelihood function used in the estimation process.

It is clearly seen in the likelihood function that there are multiple peaks present around certain angles, more specifically symmetric around $\psi = 0^\circ, 90^\circ, 180^\circ, -90^\circ$, similar to the simulation results. 
This creates challenges for the estimation process, as the \gls{pf} needs to choose the correct peak to track over time. 
It is observed that two distinct regions exist where the \gls{doa} estimate deviates from the ground truth, corresponding to intervals where the likelihood function exhibits multiple peaks. 
Firstly around 13:49:40 to 13:49:50, and secondly around 13:51:10 to 13:51:20. 
It is challenging to choose the process noise parameters for the \gls{pf} to effectively track the correct peak in these scenarios, as a higher process noise can lead to rapid switching between peaks, while a lower process noise may result in the filter getting stuck on an incorrect peak.
\begin{figure*}[tb]
    \centering
    \subfloat[$\gamma = -95$\,dBm]{
        \ifdefined\isCI
            \includegraphics{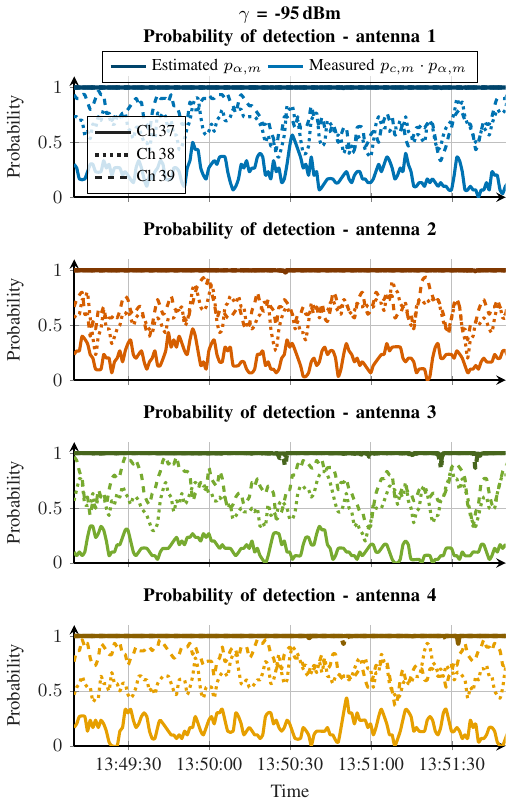}
        \else
            \tikzsetnextfilename{estimated_pd_vs_measured_95_new}
            \input{TiKz/estimated_pd_vs_measured_95_new.tex}
        \fi
        \label{fig:RealDataProbDet}
    }
    % \hspace{-2cm}
    \subfloat[$\gamma = -65$\,dBm]{
        \ifdefined\isCI
            \includegraphics{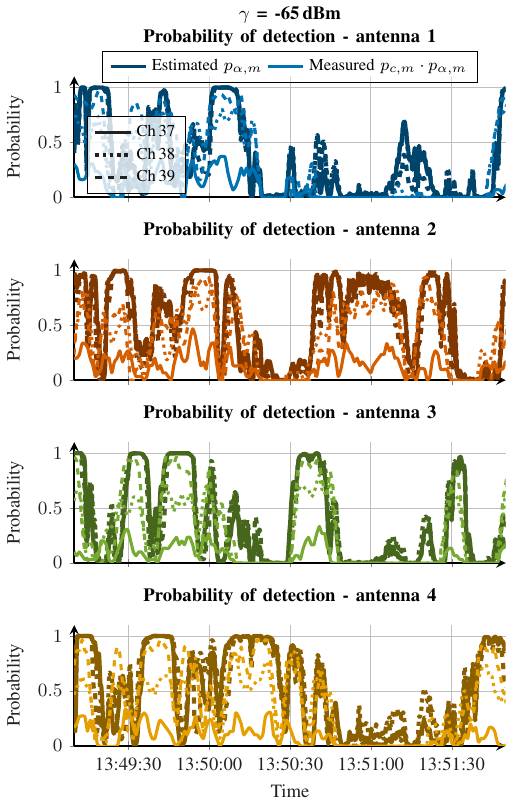}
        \else
            \tikzsetnextfilename{estimated_pd_vs_measured_65_new}
            \input{TiKz/estimated_pd_vs_measured_65_new.tex}
        \fi
        \label{fig:HighMissedDetRealDataProbDet}
    }
    \caption{The estimated probability of detection (in dark colors) and measured probability of detection (in light colors) over time for the different sensors used in the real \gls{ble} data experiments. The estimated probability of detection is estimated using the estimated $\alpha$ values and the measured \gls{gps} angle. In (\protect\subref*{fig:RealDataProbDet}) the nominal detection threshold of $\gamma = -95$\,dBm, and in (\protect\subref*{fig:HighMissedDetRealDataProbDet}) an increased detection threshold of $\gamma = -65$\,dBm.}
\end{figure*}

Looking at the estimation error in \Figref{fig:RealDataError}, the proposed method with the \gls{pf} manages to track the \gls{doa} quite accurately, even in the presence of missed detections and false negatives, with \gls{rmse} of $22.6^\circ$ over the entire experiment. 
The baseline method that does not account for missed detections performs similarly in this case, with \gls{rmse} of $24.7^\circ$, as the missed detection rates are not very high in this dataset.

\subsubsection{Increased Detection Thresholds}
If we use the estimated~$\alpha$ values from the real data experiments, and combining with the measured \gls{gps} angle, we can estimate the probability of detection for each sensor over time, as illustrated in \Figref{fig:RealDataProbDet}.

In \Figref{fig:RealDataProbDet}, the estimated probability of detection (dark lines) is generally close to 1 for all sensors, indicating that the sensors should be able to detect the signal most of the time. 
This suggests that the collected real \gls{ble} data does not enter the region of very high missed detection rates where the proposed method shows the most significant improvements in \gls{doa} estimation accuracy, as seen in the simulation results.

To further evaluate the performance of the proposed method in scenarios with higher missed detection rates, we manually adjust the threshold $\gamma$ used for detection in the real data experiments. 
By increasing the threshold, we can simulate scenarios with lower signal strengths, leading to higher missed detection rates. 
This allows us to assess the effectiveness of the proposed method in handling missed detections in more challenging conditions. 
When the threshold is increased to $\gamma = -65$\,dBm, a significant increase in missed detection rates across all sensors is observed, as illustrated in \Figref{fig:HighMissedDetRealDataProbDet}.

Here, the estimated probability of detection is significantly lower for most sensors, indicating a higher rate of missed detections. 
This creates a more challenging scenario for \gls{doa} estimation, allowing us to evaluate the performance of the proposed method under these conditions.
By applying the proposed method with false negative compensation and the \gls{pf} to the real \gls{ble} data with the increased detection threshold, the impact of higher missed detection rates on the \gls{doa} estimates can be assessed.
The results of this analysis can provide valuable insights into the effectiveness of the proposed method in handling missed detections in practical scenarios. 

In \Figref{fig:HighMissedDetRealDataResults}, we present the \gls{doa} estimates from the \gls{pf}, along with the ground truth \gls{doa} measurements from the \gls{gps} data, and the baseline method that does not account for missed detections.
The figure shows that the proposed method with the \gls{pf} is able to track the \gls{doa} more accurately compared to the baseline method, even in the presence of high missed detection rates.

The baseline method struggles to maintain accurate \gls{doa} estimates, especially in the end (from 13:51:00 and onwards), where the number of missed detections is particularly high, as shown in the colorbar in the upper plot. 
The proposed method demonstrates improved robustness to missed detections, resulting in more reliable \gls{doa} estimates.

It is worth noting that the \gls{rmse} of the \gls{doa} estimates from the proposed method in this high missed detection scenario is similar to that in the previous scenario with lower missed detection rates, while the \gls{rmse} of the baseline method is significantly worse. 
This highlights the effectiveness of the proposed method in handling missed detections, particularly in challenging scenarios with high missed detection rates.

\begin{figure}[tb]
    \centering
        \ifdefined\isCI
            \includegraphics{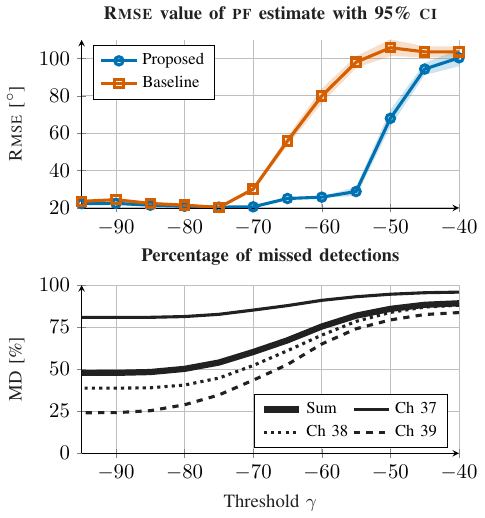}
        \else
            \tikzsetnextfilename{RMSE_vs_gamma}
            \input{TiKz/RMSE_vs_gamma.tex}
        \fi
    \caption{The \gls{rmse} of the \gls{doa} estimates from real \gls{ble} data with missed detections for different detection thresholds $\gamma$, comparing the proposed method with false negative compensation and the baseline method that does not account for missed detections. A \gls{pf} is used to track the correct peak in the likelihood function. The shaded area represents the 95\% \gls{ci} of the estimates over 10 runs with different random seeds.
    In the lower plot, the percentage of missed detections (\% MD) over all sensors is also shown as a function of the detection threshold $\gamma$.}
    \label{fig:RMSE_vs_gamma}
\end{figure}

By looking at the estimation error for different detection thresholds $\gamma$, we can further analyze the performance of the proposed method compared to the baseline method. 
In \Figref{fig:RMSE_vs_gamma}, we present the \gls{rmse} of the \gls{doa} estimates for both methods as a function of the detection threshold $\gamma$, using the same \gls{pf} settings as presented earlier. The \gls{rmse} is calculated over 10 runs with different random seeds to account for the stochastic nature of the \gls{pf}, with the shaded area representing the 95\% \gls{ci} of the estimates. 
Additionally, the percentage of missed detections (\%~MD) over all sensors is also shown as a function of the detection threshold $\gamma$ in the lower plot.
From the figure, it is observed that for low detection thresholds, up to around $\gamma = -75$\,dBm, both methods perform similarly, as the missed detection rates are relatively low in this region (around 50\% MD or lower).
However, as the detection threshold increases, leading to higher missed detection rates, the performance of the baseline method degrades significantly, with the \gls{rmse} increasing rapidly. 
In contrast, the proposed method with false negative compensation maintains a relatively stable performance, with only a slight increase in \gls{rmse} as the detection threshold increases. 
For very high detection thresholds, $\gamma = -50$\,dBm and above (around 80\% MD and higher), the proposed method starts to show a more noticeable increase in \gls{rmse}, but still performs better than the baseline method.
These results highlight the effectiveness of the proposed method in handling missed detections, particularly in challenging scenarios with high missed detection rates.

\section{Conclusions}\label{sec:conclusions}
This work demonstrates that missed detections carry valuable information that can enhance \gls{doa} estimation in systems with a known measurement model. By incorporating missed detections into a probabilistic framework, the proposed method improves robustness under challenging signal conditions and high missed detection rates.

Simulation results and real-world experiments using \gls{ble} Yagi antennas confirm improved estimation accuracy. The approach exploits both detected and missed signals, yielding more reliable \gls{doa} estimates in scenarios with few detections and low signal strength.

Despite these improvements, several limitations remain. The method is sensitive to the detection threshold, which can significantly affect performance if not properly tuned. Environmental effects such as multipath propagation and interference may degrade \gls{rssi} quality and reduce accuracy. Incorporating missed detections also increases computational complexity, which may limit real-time applicability, particularly in dynamic or low \gls{snr} settings.

As observed in simulations, the proposed method may underperform the baseline in scenarios with reduced detection efficiency due to bias introduced by missed detections within the nominal detection range, highlighting the importance of accurate detection efficiency modeling. This bias can be mitigated by increasing the number of sensors or measurements, as shown in \Secref{sec:sim_study}, or by embedding the likelihood model within a filtering framework such as the \gls{pf} in \Secref{sec:exp_study}.

Future work will first focus on relaxing the assumption of a zero false-alarm rate ($P_{FA} = 0$), since practical systems typically exhibit non-zero false alarms. Extending the model to explicitly incorporate $P_{FA}$ would enable a more complete treatment of the trade-off between missed detections and false alarms.

A second direction concerns improving robustness to non-Gaussian measurement noise. The current Gaussian assumption makes the framework sensitive to outliers arising from interference, hardware imperfections, environmental disturbances, and impulsive noise. This can be addressed through robust filtering, such as Kalman or particle filtering, with measurement gating or outlier rejection, or by adopting heavy-tailed likelihood models, such as Laplace or Student's $t$-distributions.

Further extensions include multi-source scenarios, adaptive sensor configurations, integration of complementary sensing modalities, and relaxing the assumption of persistent target existence to handle intermittent signals.
\bibliography{refs}

%(Graduate Student Member) or (Member, IEEE) or (Senior Member, IEEE) or (Fellow, IEEE)
\begin{IEEEbiography}[{\includegraphics[width=1in,height=1.25in,clip,keepaspectratio]{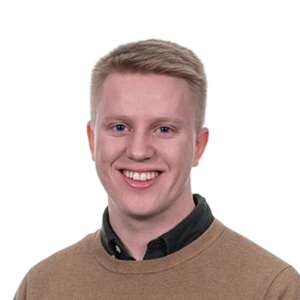}}]{Gustav Zetterqvist} received the M.Sc. degree in applied physics and electrical engineering in 2021 and the Licentiate degree in automatic control in 2024, both from Linköping University, Linköping, Sweden.

He is currently pursuing the Ph.D. degree in automatic control with the Department of Electrical Engineering, Linköping University. His research interests include signal processing, sensor fusion, and array processing with applications to acoustic, seismic, and wireless signals.

\end{IEEEbiography}

\begin{IEEEbiography}[{\includegraphics[width=1in,height=1.25in,clip,keepaspectratio]{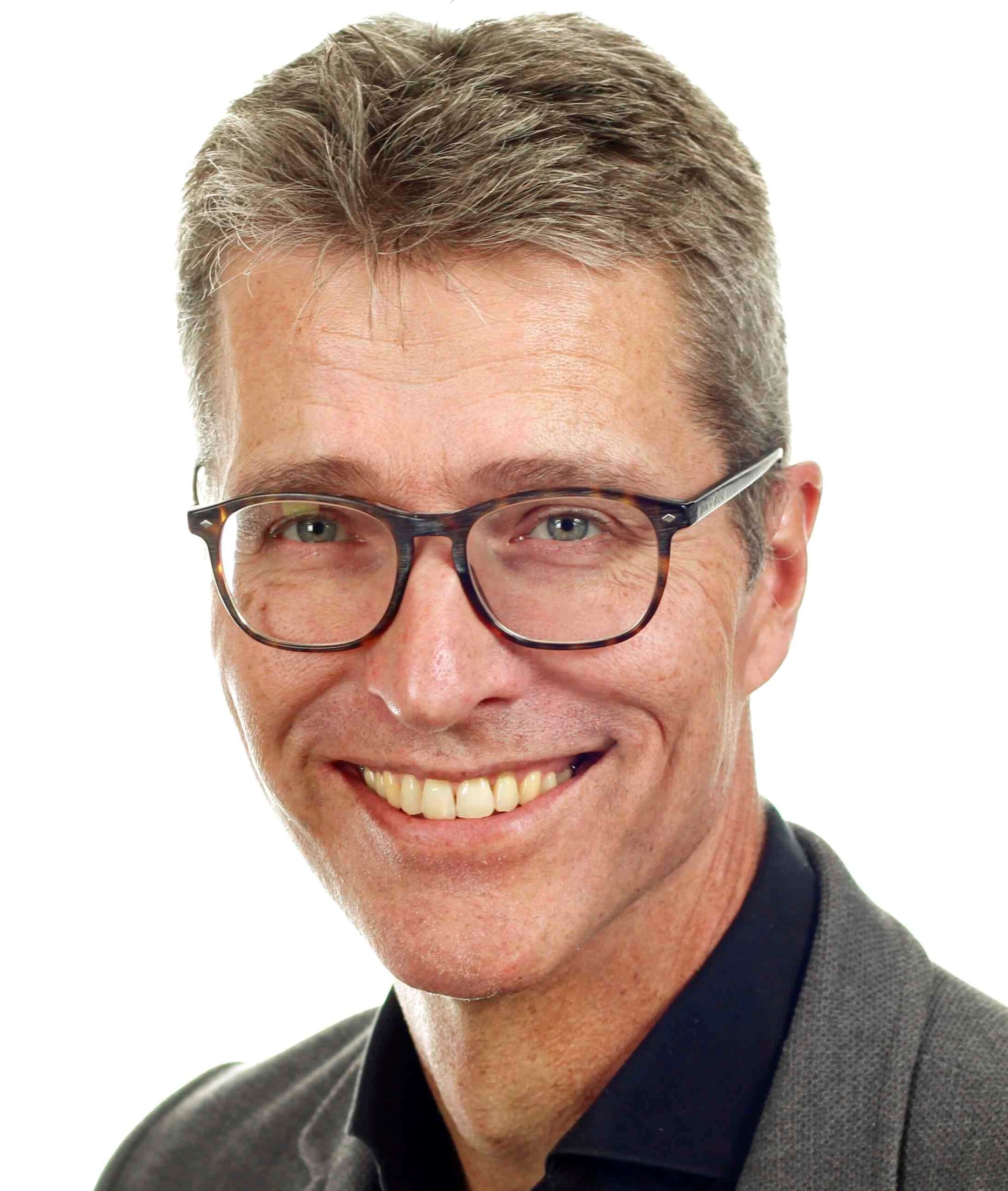}}]{Fredrik Gustafsson} (Fellow, IEEE) received the M.Sc. degree in electrical engineering in 1988 and the Ph.D. degree in automatic control in 1992, both from Linköping University. 
    
He is a Professor in Sensor Informatics with the Department of Electrical Engineering, Linköping University, since 2005. His research interests include stochastic signal processing and sensor fusion, from fundamental research to applications, where four notable spin-offs are the companies NIRA Dynamics (automotive safety systems), Softube (audio effects), Senion (indoor positioning systems) and Qulinda (animal welfare). 

He has been an Associate Editor for IEEE TRANSACTIONS ON SIGNAL PROCESSING, IEEE AEROSPACE AND ELECTRONIC SYSTEMS and EURASIP Journal on Applied Signal Processing. He was awarded the Arnberg prize by the Royal Swedish Academy of Science (KVA) in 2004, elected member of the Royal Academy of Engineering Sciences (IVA) 2007, and elevated to IEEE fellow 2011. 

He was awarded the Harry Rowe Mimno Award 2011 for the tutorial Particle Filter Theory and Practice with Positioning Applications, which was published in the AESS Magazine in 2010, he was coauthor of Smoothed State Estimates Under Abrupt Changes Using Sum-of-Norms Regularization that received the Automatica paper prize in 2014, and he was awarded the IEEE SPS Sustained Impact Paper Award 2026 for his 2001 paper Particle filters for positioning, navigation, and tracking.

\end{IEEEbiography}

\begin{IEEEbiography}[{\includegraphics[width=1in,height=1.25in,clip,keepaspectratio]{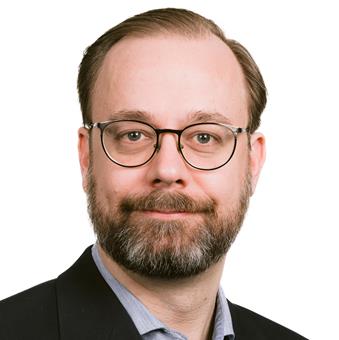}}]{Gustaf Hendeby} (Senior Member, IEEE) received the M.Sc. degree in applied physics and electrical engineering in 2002 and the Ph.D. degree in automatic control in 2008, both from Linköping University, Linköping, Sweden. 
    
He is an Associate Professor and Docent with the division of Automatic Control, Department of Electrical Engineering, Linköping University. He worked as a Senior Researcher with the German Research Center for Artificial Intelligence (DFKI) from 2009 to 2011, as a Senior Scientist with the Swedish Defense Research Agency (FOI), and held an Adjunct Associate Professor position with Linköping University from 2011 to 2015. His research interests include sensor fusion and stochastic signal processing, with applications to nonlinear problems, target tracking, and simultaneous localization and mapping (SLAM). He is the author of several published articles and conference papers in these areas. He has experience of both theoretical analysis as well as implementation aspects. 

He served as an Associate Editor for IEEE TRANSACTIONS ON AEROSPACE AND ELECTRONIC SYSTEMS in the area of target tracking and multisensor systems from 2018 to 2025, and has been a Senior Editor in the same area since 2025. Since 2024, he has served as an Associate Editor-in-Chief of the Journal of Advances in Information Fusion (JAIF). Since 2021, he has led the WASP Localization and Navigation Area Cluster. In 2022 and 2026, he served as a General Chair for the 25th and 29th IEEE International Conference on Information Fusion (FUSION), in Linköping, Sweden, and Trondheim, Norway, respectively, and as Technical Chair for the 26th, 27th, and 28th FUSION conferences in Charleston, SC; Venice, Italy; and Rio de Janeiro, Brazil, respectively. He was elected to the Board of Directors of the International Society of Information Fusion (ISIF) for the 2023-2025 and 2026-2028 terms, and currently serves as VP for Conferences.
\end{IEEEbiography}

\end{document}